\documentclass[twocolumn,letterpaper,showpacs,floatfix,aps]{revtex4}

\usepackage{graphics}
\usepackage{amssymb}

\def \beq {\begin{equation}}
\def \eeq {\end{equation}}

\begin{document}

\title{Quasinormal modes for the Vaidya metric}

 \author{ Elcio Abdalla}
 \email{eabdalla@fma.if.usp.br}
 \author{Cecilia B. M. H. Chirenti}
 \email{cecilia@fma.if.usp.br}
\affiliation{Instituto de F\'\i sica, Universidade de S\~ao Paulo\\
   C.P. 66318,  05315-970 - S\~ao Paulo, SP, Brazil.}

\author{Alberto Saa}
 \email{asaa@ime.unicamp.br}
  \affiliation{Departamento de Matem\'atica Aplicada, UNICAMP \\
   C.P. 6065,  13083-859 - Campinas, SP, Brazil.}


\begin{abstract}
We consider here scalar and electromagnetic perturbations for the Vaidya metric
in double-null coordinates. Such an approach allows one to go a step further in the
analysis of quasinormal modes for time-dependent spacetimes. Some recent results
are refined, and a new non-stationary behavior corresponding to some sort of   
inertia for quasinormal modes is identified. Our conclusions can enlighten some aspects
of the wave scattering by black holes undergoing  some mass accretion processes.
\end{abstract}
\maketitle

\section{Introduction}

The quasinormal modes analysis is a paradigm for the study of gravitational
excitations of black holes. For a comprehensive review on the vast previous literature,
see \cite{Nellert-99,Kokkotas-99}. 
We stress that  the quasinormal modes (QNM) of black holes are
also experimentally relevant  since they could, in principle, be detected in gravitational
waves detector such as LISA (see, for instance, \cite{berti05} and references therein).
The robustness of  QNM  analysis
has been established by considering several kinds of generalizations, as, for 
instance, relaxing the asymptotic flatness condition\cite{Brady-97,Molina-04,Du-04}
or considering time-dependent situations\cite{Hod-02,Xue-04,Cheng-04}. The case
of asymptotic Anti-de-Sitter black holes has deserved special attention due
to its relevance to the AdS/CFT conjecture\cite{Chan-97,Cardoso-124015,
Horowitz-00,Wang-00,Wang-01,Cardoso-084017,Berti-03,Konoplya-02,
Birmingham-02,Zhu-01,Musiri-03,Cardoso-084020,Cardoso-03}. 

The analysis of QNM for time-dependent situations is of special physical interest  since
it is expect  that a black hole can enlarge its mass by some accretion
process, or even lose   mass by some other process, including Hawking radiation.
The late time tail for a Klein Gordon field under a time-dependent potential was
first considered in \cite{Hod-02}, where some influence of the temporal dependence 
of the potential over the characteristic decaying tails was reported.
Such a kind of time-dependent potential arises naturally when the Vaidya metric
is considered\cite{Xue-04,Cheng-04}.
The Vaidya metric, which 
in radiation coordinates $(w,r,\theta,\phi)$
has the form
\beq
\label{Vaidya}
ds^2 = -\left(1-\frac{2m(w)}{r}\right)dw^2+2cdrdw + r^2d\Omega^2,
\eeq
where $d\Omega^2 = d\theta^2 + \sin^2\theta d\phi^2$, 
$c=\pm 1$, is a solution
of Einstein's equations with spherical symmetry in the eikonal
approximation to a radial flow of unpolarized 
radiation. For the case of an ingoing radial flow, $c=1$ and
$m(w)$ is a monotone increasing mass function in the advanced
time $w$, while $c=-1$ corresponds to an outgoing radial flow,
with $m(w)$ being in this case 
a monotone decreasing mass function in the retarded
time $w$. The metric (\ref{Vaidya}) is the starting point for the
QNM analysis of varying mass black-holes. 

In \cite{Cheng-04}, massless
scalar fields were studied on an electrically charged version of the
Vaidya metric. Basically, two kinds of continuously varying mass functions were
considered:
\begin{equation}
\label{eq26}
 m(w) = \left\{ {{\begin{array}{l l}
 {m_1 } & {w \le w_1 }, \\
 {m_1 } f(w) & {w_1 \le w \le w_2 }, \\
 {m_2 } & {w \ge w_2 }, \\
\end{array} }} \right. \\
\end{equation}
with $f(w)=(1-\lambda w)$ and $f(w)=\exp{(-\alpha w)}$,   called, respectively,
linear and  
exponential models. Generalized tortoise coordinates were introduced,
and a standard numerical analysis was done. The conclusion was that, as a
first approximation, the QNM for a stationary Reissner-Nordstr\"om black
hole with mass $m(w)$ and charge $q(w) = q_0 m(w)$ are still valid, as if some stationary adiabatic regime was indeed
governing the QNM dynamics for time-dependent spacetimes. In principle, one should not
expect such stationary behavior for very rapid varying mass functions,
however the analysis of \cite{Cheng-04} was not able to identify any breaking
of such an adiabatic regime. We notice also that both linear and exponential models,
inspired by some known exact solutions\cite{WL} for which generalized
tortoise coordinates could be explicitly constructed, would hardly correspond to
physically realistic cases. Both cases have $C^0$-class mass functions,
implying the existence of some infinitesimal shell distributions 
of matters for $w=w_0$ and
$w=w_1$ whose interpretation and role are still unclear.

We attack here the QNM problem by considering the Vaidya metric in double-null
coordinates\cite{GS}. The main advantage of such an approach 
is the possibility of considering any (monotone) mass function, allowing, for instance,
the analysis of smooth mass functions  that could correspond  to physically more
relevant situations, free
of obscure infinitesimal matter shells. Furthermore, the use of double-null 
coordinates has improved considerably the overall 
precision of the numerical analysis, allowing us to identify the breakdown
of the adiabatic regime, characterized by the appearing of a non-stationary inertial
effect for the QNM in the case of rapid varying mass functions. Once the
stationary regime is reached, the standard QNM results hold, reinforcing,
once more, the robustness of the QNM analysis. Our results can be used as a first
approximation to describe the propagation of small perturbations around 
astrophysically realistic 
situations where black holes undergo some mass accretion  processes.
We notice that the non-stationary inertial effect presented here is in
perfect agreement with the non-linear analysis preformed in \cite{Zlochower:2003yh},
which reported extra redshift effects on the QNM frequencies due to the growth
of the effective black hole mass.

We organize this paper as
follows. The next section is devoted to a brief introduction to the semi-analytical
approach\cite{GS} for the Vaidya metric in double-null coordinates. Section
III presents the main issues and results of our numerical analysis. The 
last section is left to some closing remarks.

\section{The Vaidya metric in double-null coordinates}

Double-null coordinates are specially suitable   for time-evolution problems
as the QNM analysis. However, the difficulties of constructing double-null coordinates
for non-stationary spacetime are well known. The Vaidya metric is a typical 
example (see \cite{GS} and references therein). Indeed, it is known that the
problem of constructing  double-null coordinates for generic mass functions
is not analytical soluble in general\cite{WL}. The semi-analytical approach proposed in
\cite{GS} is the starting point for our analysis here. It consists, basically, in
considering the Vaidya metric in double-null coordinates {\em ab initio},
avoiding the need of constructing any coordinate transformation.
The spherically symmetric line element in double-null coordinates is
\beq
\label{uv}
ds^2 = -2f(u,v)du\,dv + r^2(u,v)d\Omega^2,
\eeq
where $f(u,v)$ and $r(u,v)$ are smooth non vanishing functions.
The energy-momentum tensor
of a unidirectional radial flow of unpolarized radiation  
in the eikonal approximation is 
given by
\beq
\label{T}
T_{ab} = \frac{1}{8\pi}h(u,v)k_a k_b,
\eeq
where $k_a$ is a radial null vector. 
The Einstein's equations for the case of a flow in the $v$-direction
can then be reduced to the following set of equations\cite{WL,GS}
\begin{eqnarray}
\label{ef}
f(u,v) &=& 2B(v)\partial_u r(u,v), \\
\label{er}
\partial_v r(u,v) &=& -B(v)\left( 1-\frac{2m(v)}{r(u,v)}\right), \\
\label{eh}
h(u,v) &=& -4\frac{B(v) m'(v)}{r^2(u,v)},
\end{eqnarray}
where $B(v)$ and $m(v)$ are arbitrary functions obeying, according
to the weak energy
condition,   
\beq
\label{WE}
B(v)m'(v) \le 0,
\eeq
where the prime denotes the derivative with respect to $v$.
The solution of (\ref{ef})-(\ref{eh}) will correspond to the Vaidya metric
in double-null coordinates, as one can interpret from (\ref{uv}) and (\ref{T}).
For $m'(v)\ne 0$, the choice 
\beq
B=-\frac{1}{2}{\rm sign}(m') 
\eeq 
allows one to interpret $m(v)$ as the mass
of the solution and $v$ as the proper time as measured in the rest frame at infinity
for the asymptotically flat case\cite{WL,GS}.
Note that if the weak energy condition (\ref{WE}) holds, 
the function $m(v)$ is monotone, implying that 
the radial flow must be ingoing or outgoing for all $v$. 
It is not possible, for instance, 
to have ``oscillating'' mass functions $m(v)$. 

The semi-analytical approach of \cite{GS} consists in a strategy to construct numerically
the functions $f(u,v)$, $r(u,v)$ and $h(u,v)$ from the Eq. (\ref{ef})-(\ref{eh}), and to infer
the underlying causal structure. 
Eq. (\ref{er}) along $u$ constant is a first order ordinary differential
equation in $v$. One can evaluate the function $r(u,v)$ in
any point by solving the $v$-initial value problem knowing
$r(u,v_0)$. The trivial example of Minkowski spacetime ($m=0$),
for instance, can be obtained\cite{GS} by choosing $r(u,0)=u/2$. Once we have $r(u,v)$,
we can evaluate $f(u,v)$ and $h(u,v)$ from (\ref{ef}) and (\ref{eh}).
Further details of the method can be found in \cite{GS}. We consider here the
following smooth mass function
\begin{equation}
m(v) = m_{1}+\frac{m_{2}-m_{1}}{2}\left[1+\tanh \rho(v-v_{1})\right]
\label{hyperbolic}
\end{equation}
where $\rho$ and $v_{1}$ are also constant parameters. 
For sake of comparison with the results of \cite{Cheng-04}, we also consider a linear model.
Fig. 1 depicts the causal structure corresponding to the hyperbolic mass
\begin{figure}[ht]
\resizebox{\linewidth}{!}{\includegraphics*{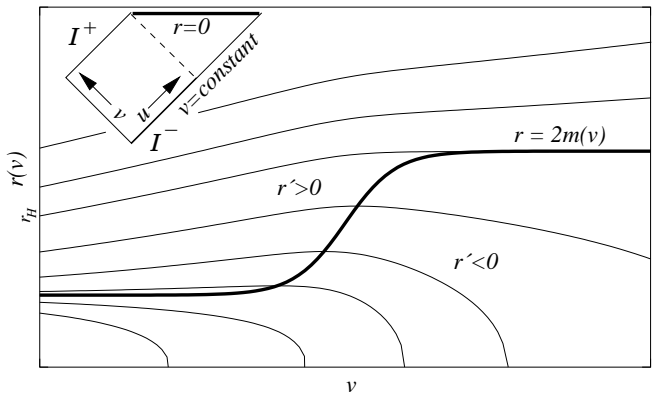}}
\caption{Lines of constant $u$ for the solution of (\ref{er}) with
the hyperbolic mass function (\ref{hyperbolic}), with $m_2>m_1$ and $\rho>0$. 
All solutions   in the region below the line 
$r=2m(v)$ (the apparent horizon) have $r'<0$,
where the prime denotes the derivative with respect to $v$.
Any solution that enters into this region will reach
the singularity at $r=0$ with finite $v$. On the other hand, solutions
confined to the $r'>0$ region always escape from the singularity and
reach ${\cal I^+}$. In the present case, 
there 
exists an event horizon (the dashed line on the inserted conformal
diagram) close to  the solutions $r_{\rm H}(v)$.
}
\end{figure} 
 function (\ref{hyperbolic}) obtained from the semi-analytical approach. For positive $\rho$
 and $m_2>m_1$,
 for instance, it represents a black hole with mass $m_1$ receiving a radial flux of radiation
 and, consequently, enlarging continuously its mass until reaching $m_2$.
The choice of the initial condition for solving (\ref{er}) is a rather subtle issue\cite{GS}.
For our purposes here, we note only that demanding $\partial_u r(u,v_0)\ne 0$ is sufficient
to guarantee that, for any $m(v)$, the underlying spacetime causal 
structure does not depend on the initial 
condition $r(u,v_0)$. For constant $m$  and $B=-1/2$, Eq. (\ref{er}) can be easily integrated,
leading to
\begin{equation}
\label{ic}
r(u,v) + 2m \ln|r(u,v)-2m| - \frac{v}{2} = P(u),
\end{equation}
where $P(u)$ is an arbitrary function of $u$.
It is also possible to solve Eq. (\ref{er}) analytically for the linear and exponential mass 
functions\cite{WL}. Eq. (\ref{ic}) is our reference to choose the initial conditions $r(u,v_0)$.

\section{Scalar and electromagnetic perturbations}

In the coordinate system (\ref{uv}), for any desired form of 
the mass function $m(v)$, the equations for scalar and electromagnetic 
perturbations can be easily put  in the form\cite{Kokkotas-99}
\begin{equation}
\frac{\partial^{2}\psi}{\partial u\partial v}  +
V(u,v)f(u,v)\psi = 0,
\label{scalar}
\end{equation}
where the potential $V(u,v)$ is given by
\begin{equation}
V(u,v) = \frac{\ell(\ell + 1)}{2r^{2}(u,v)} +
\sigma \frac{ m(v)}{r^{3}(u,v)},
\end{equation}
where $\sigma=1$ and $\sigma=0$ correspond, respectively, to the scalar and  to the 
electromagnetic cases. For a given mass function $m(v)$, one first evaluates the function
$f(u,v)$ and $r(u,v)$ with the semi-analytical approach of the previous section, and then
the characteristic problem corresponding to (\ref{scalar}) can be solved with the usual second order
characteristic algorithm\cite{Kokkotas-99}, where the initial data are specified
along the two null surfaces $u = u_{0}$ and $v = v_{0}$. Since the basic
aspects of the field decay are independent of the initial conditions
(this fact is confirmed by our simulations),
we use the Gaussian initial condition
\begin{equation}
\psi(u=u_0,v) = \exp\left[-\frac{(v - v_c)^2}{2\sigma^2}\right] \ ,
\end{equation}
and $\psi(u, v=v_0) = \rm const$. Our typical numerical grid is large enough to
assure that we can set effectively this last constant to zero.
After the integration is completed, the values $\psi(u_{max}, v)$ 
are extracted,  where  $u_{max}$ 
is  the  maximum value  of $u$  on  the numerical grid. Taking
sufficiently large $u_{max}$, we have good
\begin{figure}[ht]
\resizebox{\linewidth}{!}{\includegraphics*{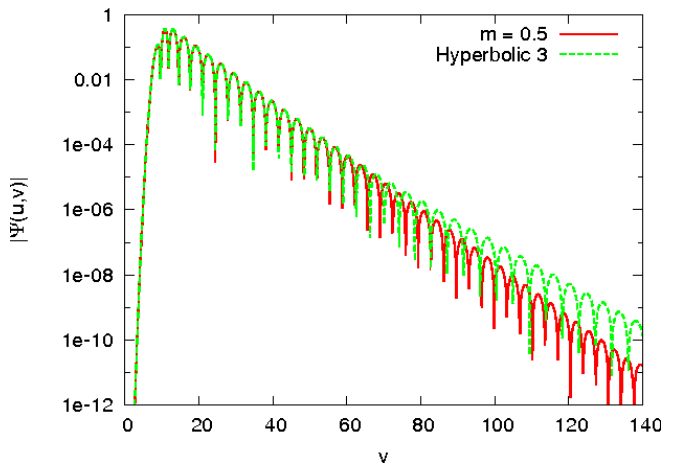}}
\caption{Values of the electromagnetic perturbations $\psi(u_{max}, v)$ with
$\ell=2$ for a Schwarzschild
black hole with mass $m=0.5$ and for the Vaidya metric with 
the hyperbolic mass function
(\ref{hyperbolic}), where $v_1=75$, $\rho = 0.08$, $m_1=0.5$, and $m_2=0.65$ (Hyperbolic 3). One can clearly
appreciate, for the time-dependent case, the slowing down of the oscillation frequency and damping 
taking place after $v_1$.}
\end{figure}
approximations for the wave function at the event 
horizon. Figure 2 presents an example of $\psi(u_{max}, v)$ data 
for a hyperbolic increasing mass function and its comparison with the case of 
a Schwarzschild
black hole. Similar results hold if we extract the data for other values of $u$.
All the analysis presented here corresponds to the data extracted on the horizon $u_{max}$.
Note that the choice of $u_{max}$ is a matter of numerical convenience. Since 
the QNM corresponds to eigenstates of an effective Schroedinger 
equation\cite{Kokkotas-99}, the associated complex eigenvalues can be read out anywhere once
the asymptotic regime is attained.

From the function $\psi(u_{max}, v)$,
one can infer the characteristic frequencies $\omega$ of the 
damped oscillating modes. Since 
   the frequencies $\omega$ are themselves time-dependent (see Fig. 2), 
for a given $v$, $\omega(v)$ is determined locally by a non-linear $\chi^2$-fitting
by using some damped cycles of $\psi$ around $v$.
\begin{figure}[ht]
\resizebox{\linewidth}{!}{\includegraphics*{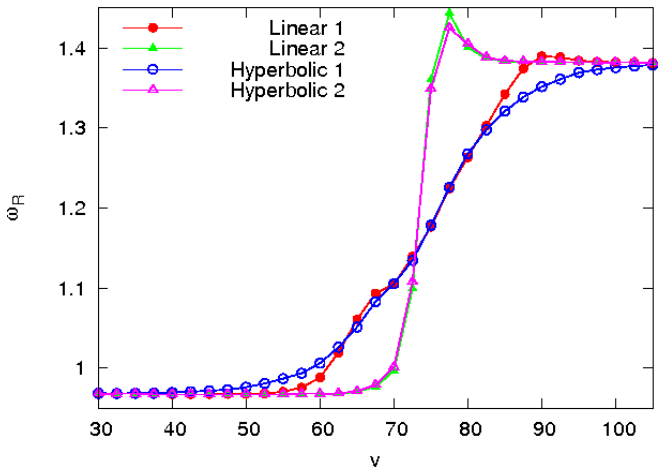}}
\caption{The real   ($\omega_{\rm R}$)    part of the frequency for scalar perturbations
with
$\ell=2$ as function of $v$
for decreasing linear and hyperbolic mass functions. For all the cases,
$m_1=0.5$ and $m_2=0.35$. 
Linear 1: $v_1=60$, $v_2=90$. 
Linear 2: $v_1=74.5$, $v_2=75.5$.
Hyperbolic 1: $v_1=75$, $\rho=0.08$.
Hyperbolic 2: $v_1=75$, $\rho=0.8$. The imaginary part $\omega_{\rm I}$ has similar behavior as
for the increasing mass case (Fig. 4), see the text.
}
\end{figure}
We perform an exhaustive numerical analysis for many different mass functions
$m(v)$. Figures 3 and 4 present the real  ($\omega_{\rm R}$)  and imaginary
($\omega_{\rm I}$)
part of the frequencies of
the perturbations as
function of $v$. They correspond, respectively, to some decreasing and increasing  mass functions, 
and are typical for all values of $\ell$, all kinds of perturbations (scalar and electromagnetic) and
different initial conditions.
The imaginary part $\omega_{\rm I}$ exhibits similar time-dependence, although its 
values are typically known with lower precision than $\omega_{\rm R}$. 
This is not, however, due to numerical errors in the integration procedure. In fact, the convergence of the
second order characteristic algorithm is quite good. For instance, the differences
in the calculated frequencies of Fig. 3 and 4 obtained by using
as integration steps $\Delta u = \Delta v = 0.2, 0.1,$ and 0.05, are smaller than
the size of the points used in the figures, and no difference is detected for even
smaller steps.
We credit the irregularities 
observed for $\omega_{\rm I}$ to the local $\chi^2$-fitting described   above.
For subcritical damped oscillations as the QNM considered here, the oscillation
frequencies $\omega_{\rm R}$ can be easily determined from a few damped cycles,
while  for the damping term $\omega_{\rm I}$ one typically needs many more cycles
in order to get a similar precision. By taking a large number of cycles we 
will tend to smear the calculated frequencies, leading to some kind of average  
and not to the desired instantaneous values. We opt, then, to take as few as possible
cycles, paying the price of having some irregularities for $\omega_{\rm I}$.

Let us take for a close analysis,
for instance, the case of decreasing $m(v)$  (Fig. 3). 
For the rapid varying cases (Linear 2 and Hyperbolic 2), one clearly sees the inertial
effect for $\omega_{\rm R}$ near $v=75$. The function $\omega_{\rm R}(v)$ does not
follow the track corresponding to $m^{-1}(v)$, as one would expect for a stationary adiabatic
regime, and as it really does for the Hyperbolic 1 case. After the rapid
increasing phase, $\omega_{\rm R}$ acts as it would have some intrinsic inertia, reaching
a maximum value that is bigger than $\omega_{\rm R}(\infty)$, implying the relaxation
corresponding to the region with $\omega'_{\rm R}(v)<0$ for $v>75$. We also notice
that, for the rapid varying case, one could not detect sensible differences between
the smooth hyperbolic case and the $C^0$ linear one. Nevertheless, one can see that the slower
linear case (Linear 1) exhibits some inertial effects close the the second matching
point of $m(v)$. In all other regions, it follows the track corresponding to $m^{-1}(v)$. 
No appreciable difference  in the transients of scalar and electromagnetic perturbations
was detected.
As we have already mentioned, such transient inertial behavior passed unnoticed by the QNM analysis in
radiation coordinates performed in \cite{{Cheng-04}}.
Analogous conclusions hold for the case of increasing $m(v)$  (Fig. 4).
\begin{figure}[ht]
\resizebox{\linewidth}{!}{\includegraphics*{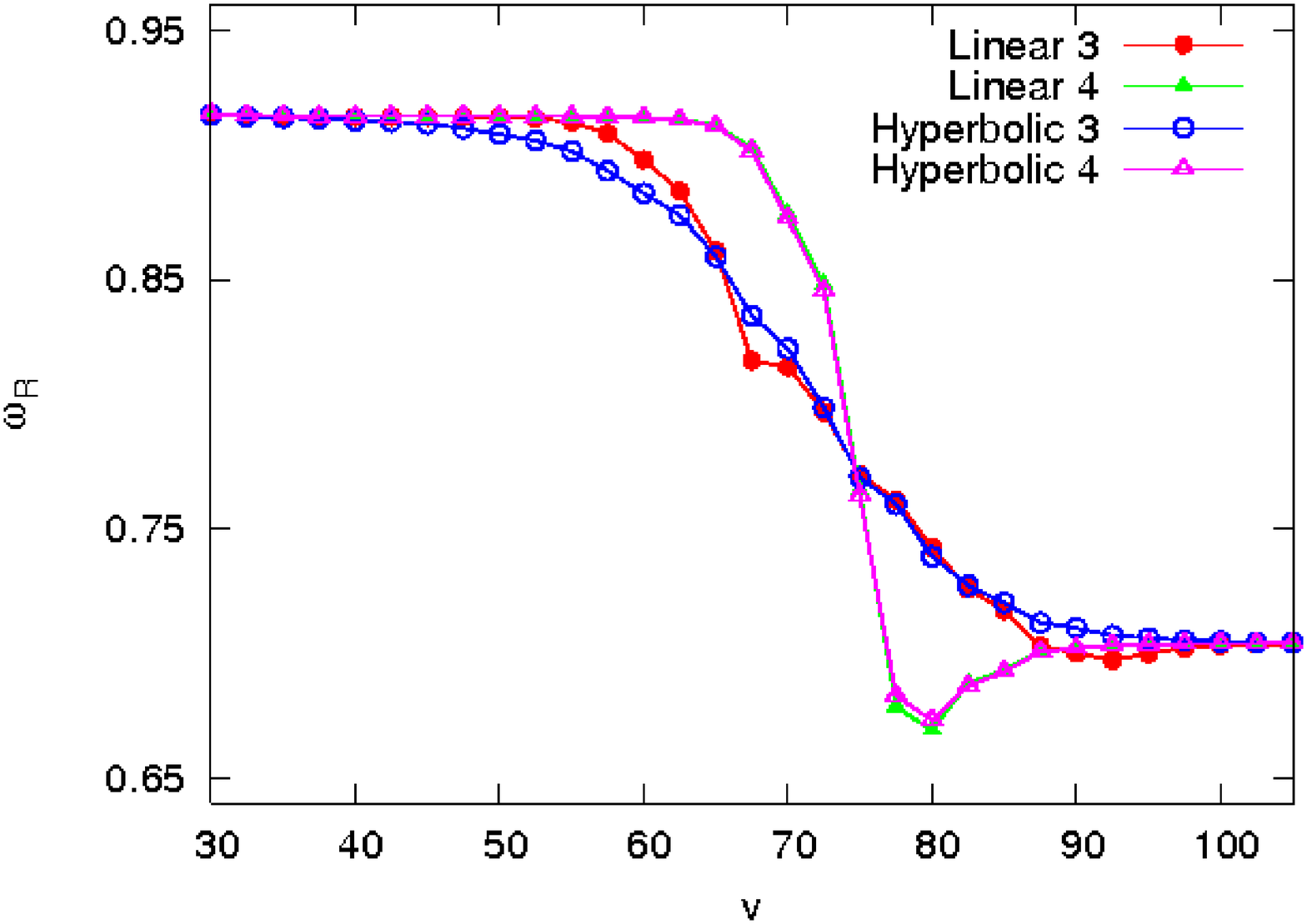}}
\resizebox{\linewidth}{!}{\rotatebox{270}{\includegraphics*{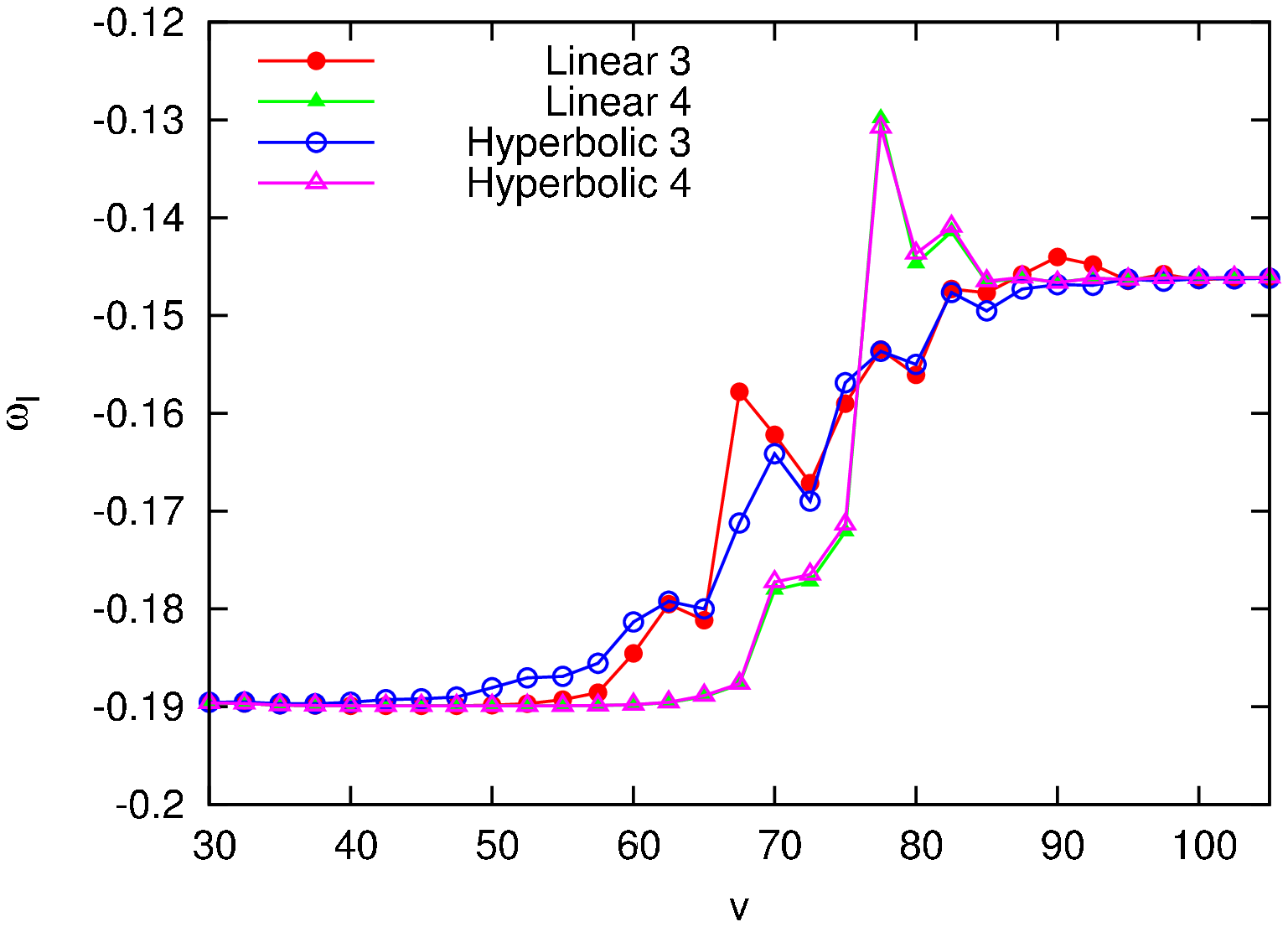}}}
\caption{The real   ($\omega_{\rm R}$)  and imaginary ($\omega_{\rm I}$)  part of the frequency for electromagnetic perturbations
with
$\ell=2$ as function of $v$
for increasing linear and hyperbolic mass functions. For all the cases,
$m_1=0.5$ and $m_2=0.65$. 
Linear 3: $v_1=60$, $v_2=90$. 
Linear 4: $v_1=74.5$, $v_2=75.5$.
Hyperbolic 3: $v_1=75$, $\rho=0.08$.
Hyperbolic 4: $v_1=75$, $\rho=0.8$. 
}
\end{figure}
In fact, since the increasing mass functions considered here can be obtained
from the decreasing ones by a time-reversal $m(v)\rightarrow m(-v)$, the underlying
causal structures are equivalent\cite{GS}, and
the corresponding
QNM are also related. Under a time-reversal, the graph of $\omega_{\rm R}$ (and of 
$\omega_{\rm I}$)   
must be  reflected on a horizontal axis. 

We could infer from our numerical simulations the situation corresponding to
the onset of the QNM non-stationary inertial behavior. The deviation from the stationary 
regime is measured by the second derivative of the mass function $m''(v)$.
Heuristically, one should expect the appearing of non-stationary behavior when $|1/m''|$
is smaller than a certain characteristic time of the system, what should 
prevent the  system to relax into an adiabatic regime. There are two
characteristic times associated to the QNM of black holes: the oscillation
period $2\pi/\omega_{\rm R}$ and the damping time $|1/\omega_{\rm I}|$. 
The onset of the inertial behavior is associated to the second one. We verify appreciable
deviations from the stationary regime whenever $|m''(v)|$ is of the same order (or larger)
than $|\omega_{\rm I}|$. For instance, for Hyperbolic 2 and Hyperbolic 4 data set  
(corresponding to the inertial behavior depicted in Figs. 3 and 4) we 
have $|m''_{\rm max}/\omega_{\rm I}^{\rm final}| \approx 25\%$,
while for Hyperbolic 1 and Hyperbolic 3 (the stationary behavior) such ratio is 
100 times smaller. Estimating   the magnitude of the inertial effect from our
numerical simulations without an approximated analytical model seems to be much harder.
Again heuristically, one expects that the magnitude of the effect be proportional
to $|m''|\tau$, where $\tau$ is the time interval along which $|m''|\gtrapprox|\omega_{\rm I}|$.
We could check  that the relative variations in frequencies during
the inertial behavior for the smooth case are always limited by the ratio $|m''/\omega_{\rm I} |$.
Further analytical work is certainly necessary to enlighten this point. We finish
by noticing that for the $C^0$ linear case, $|m''_{\rm max}|$ is always large,
increasing indeed with $1/\Delta v$, and hence non-stationary inertial behavior is reported in all
simulations using $C^0$ linear data set. Since $\tau$ is very small in such cases
(of the same order of $\Delta v$), the magnitude of the inertial effects are
also typically small.

\section{Final remarks}

All situations we considered in the present work involve mass functions 
corresponding to an initial black hole with mass $m_1$ undergoing some
accretion/losing mass process and ending with a mass $m_2$. Such an
``asymptotic Schwarzschild''
choice assures us that the spacetime has the usual black hole causal
structure for $v\rightarrow\pm\infty$ and, consequently, that QNM can be
defined in the usual way and that the corresponding frequencies can be
properly compared. For all the cases considered, the inertial transients
dissipate away and $\omega(v)$ tends to follow the track of
$m^{-1}(v)$ rather quickly, rendering  the robustness of the numerical QNM analysis. 
The integration in double-null coordinates has turned out to be much more efficient 
than the integration in radiation coordinates\cite{Cheng-04}, allowing us to 
reach the precision necessary to unveil the reported non-stationary behavior
with quite modest computational resources. Despite the radiation coordinates are
known to be defective at the horizon\cite{radhorizon}, we believe that numerical
analysis as those one presented in \cite{Cheng-04} are still confident since
 the QNM analysis   is always concerned to the exterior region of the
black hole.
However, as confirmed by the present calculations,   algorithms based on the
characteristic integration are typically by far more efficient.

An interesting extension of this work would be analysis of the highly damped
QNM (overtones, $n>0$). Since for such overtones the ratio $|\omega_{\rm I}/\omega_{\rm R}|$
is always larger than for the $n=0$ QNM considered here, including, for sufficient
large $n$, cases for which $|\omega_{\rm I}/\omega_{\rm R}|>1$, it would be interesting
to check if the non-stationary inertial behavior could somehow be attenuated for $n>0$.
We notice that the
numerical analysis presented here cannot be extended directly to the  
  $n>0$ case since one cannot identify the $n>0$ frequencies with sufficient
accuracy. We believe this could be attained, in principle, by means of the WKB
approximation.

Although the typical astrophysical situations of mass accretion  for   black holes
 will hardly keep spherical symmetry intact during intermediate stages, 
 our results can be used as a first approximation for the scattering by this sources
 since, as it is well known, after
 the transient phases, the system should accommodate itself in a stationary spherical symmetric
 configuration. One must, however,
 keep in mind that non-stationary inertial behavior
 in the QNM frequencies is expected to take place whenever $|m''|\gtrapprox|\omega_{\rm I}|$, 
 rendering the analysis of the rapid variating situations
 a   subtler task.

 The evaporation by Hawking radiation could indeed be considered as a physically
 genuine process where the black hole mass decreases and spherical symmetry is maintained.
Our approach can be applied for this case, with some crucial remarks about the causal structure
of the spacetime left behind after the evaporation\cite{ACS}.

We finish by noticing, as it is well known, that the damped oscillations correspond to
an intermediate phase in the wave scattering by asymptotic flat black holes. The very last
final phase corresponds, indeed,   to a power law decay. In the problems considered here, the power
law tail typically appears for large values of $v$ for which the QNM already settled down
in the stationary phase, with no trace from the transients. We do not detect, in this case,
any influence of the time-dependence of the potential in these tails. There is no contradiction
with the results reported in \cite{Hod-02}. This is a consequence of our choice of 
``asymptotic Schwarzschild'' mass functions, for which the corresponding potential has
no resemblance with those ones considered in \cite{Hod-02}.

\acknowledgements

The author are grateful to FAPESP and CNPq for the financial support.

\end{document}